\documentclass{elsart3}
\usepackage{graphics}
\usepackage{epsfig}
\usepackage{amssymb}
\usepackage{cite}
\usepackage{subfigure}

\begin{document}
  \newcommand{\greeksym}[1]{{\usefont{U}{psy}{m}{n}#1}}
  \newcommand{\umu}{\mbox{\greeksym{m}}}
  \newcommand{\udelta}{\mbox{\greeksym{d}}}
  \newcommand{\uDelta}{\mbox{\greeksym{D}}}
  \newcommand{\uOmega}{\mbox{\greeksym{W}}}
  \newcommand{\uPi}{\mbox{\greeksym{P}}}
  \newcommand{\ualpha}{\mbox{\greeksym{a}}}
  \begin{frontmatter}


\vspace*{-11mm}{\it \begin{flushleft} \small
Talk presented at the 14$^{th}$ International Workshop on Vertex Detectors (VERTEX2005),\\
November 7-11 2005, Chuzenji Lake, Nikko, Japan.
\end{flushleft}}
\vspace{-1.5cm}
\title{Sensor simulation and position calibration \\ for the CMS pixel detector}

\author[uniz]{V.~Chiochia\corauthref{cor1}}, \ead{vincenzo.chiochia@cern.ch}
\author[uniz]{E.~Alag\"oz},
\author[jhu]{M.~Swartz}

\corauth[cor1]{Corresponding author}
\address[uniz]{Physik Institut der Universit\"at Z\"urich-Irchel, 8057 Z\"urich, Switzerland}
\address[jhu]{Johns Hopkins University, Baltimore, MD 21218, USA}
\begin{abstract}
In this paper a detailed simulation of irradiated pixel sensors
was used to investigate the effects of radiation damage on charge sharing and
position determination.
The simulation implements a model of radiation damage by including two
defect levels with opposite charge states and trapping of charge carriers.
We show that charge sharing functions extracted from the simulation
can be parameterized as a function of the inter-pixel position and used to
improve the position determination. For sensors irradiated to $\Phi=5.9\times10^{14}$ n$_{\rm{eq}}/$cm$^2$
a position resolution below 15 $\mu$m can be achieved after calibration.
\end{abstract}

\end{frontmatter}

\section{Introduction}
The CMS experiment, currently under construction at the Large Hadron Collider
(LHC) will include a silicon pixel detector~\cite{CMSTrackerTDR:1998} to allow tracking in the region closest
to the interaction point. The detector will be a key component for reconstructing interaction
vertices and heavy quark decays in a particulary harsh environment, characterized by
a high track multiplicity and heavy irradiation.
At the full LHC luminosity the innermost layer, with a radius of 4.3 cm, 
will be exposed to a particle fluence of $3\times10^{14}$ n$_{\rm{eq}}/$cm$^2$/yr.
                                                      
In order to evaluate the effects of irradiation and optimize the algorithms
for the position determination a detailed  simulation of the pixel sensors
was implemented.
In~\cite{Chiochia:2004qh} we have proven
that it is possible to adequately describe
the charge collection characteristics of heavily irradiated silicon pixel detectors in terms
of a tuned double junction model which produces a doubly peaked electric field profile across the sensor.  
The modeling is supported by the evidence of doubly peaked electric fields obtained from beam
test measurements presented in~\cite{Dorokhov:2004xk}.  
The dependence of the modeled trap concentrations upon fluence was presented in~\cite{Chiochia:2005ag}
and the temperature dependence of the model was discussed in~\cite{Swartz:2005vp}.
Charge sharing and hit reconstruction after irradiation was investigated
in~\cite{Chiochia:TIME05} using the so called ``$\eta$-technique''. In this paper we present an
alternative approach to hit reconstruction and calibration based on
charge sharing functions extracted from the sensor simulation.   

This paper is organized as follows: The sensor simulation is described
in Section~\ref{sec:sensor_simulation}, in Section~\ref{sec:highlevelcalib} 
the hit reconstruction and calibration is discussed. The
conclusions are given in Section~\ref{sec:conclusions}.

\section{Sensor simulation~\label{sec:sensor_simulation}}

The results presented in this paper rely upon a detailed 
sensor simulation that includes the modeling of irradiation effects in silicon.
The simulation, {\sc pixelav}~\cite{Swartz:2003ch,Swartz:CMSNote,Chiochia:2004qh}, 
incorporates the following elements: an accurate model of charge deposition by primary hadronic
tracks (in particular to model delta rays); a realistic 3-D intra-pixel electric field map; 
an established model of charge drift physics including mobilities, Hall Effect, and 3-D diffusion; 
a simulation of charge trapping and the signal induced from trapped charge; and a 
simulation of electronic noise, response, and threshold effects.  
The intra-pixel electric field map was generated using {\sc tcad 9.0} \cite{synopsys} to simultaneously 
solve Poisson's Equation, the carrier continuity equations, and various charge transport models.  

The simulated devices correspond to the baseline sensor design for the CMS
barrel pixel detector. The sensors are ``n-in-n'' devices, designed to collect
charge from n$^+$ structures implanted into n- bulk silicon.
The simulated samples were 22x32 arrays of 100x150 $\mu$m$^2$ pixels. 
The substrate was 285 $\mu$m thick, n-doped silicon. The donor concentration
was set to $1.2\times10^{12}$ cm$^{-3}$ corresponding to a depletion
voltage of about 75 V for an unirradiated device. 
The 4 T magnetic field was set as in the CMS configuration 
and the sensor temperature to -10$^\circ$ C.
The simulation did not include the ``punch-through'' structure on the n$^+$ implants which is used
to provide a high resistance connection to ground and to provide the possibility
of on-wafer IV measurements. 

The effect of irradiation was implemented in the {\sc tcad} simulation by including
two defect levels in the forbidden silicon bandgap with opposite
charge states and trapping of charge carriers. The activation energies
of the donor and acceptor traps were set to $(E_{V}+0.48)$ eV and $(E_{C}-0.525)$
eV, respectively, where $E_{V}$ and $E_{C}$ are the valence and conduction
band energy level, respectively~\cite{Eremin:2002wq}. 
The trap densities and the capture cross sections 
for electrons and holes were obtained by fitting the model to beam test data 
as described in~\cite{Chiochia:2004qh,Chiochia:2005ag}. The simulated irradiation fluence
was $\Phi = 5.9 \times 10^{14}$ n$_{\rm{eq}}$/cm$^2$
and the reverse bias was set to 300 V.

\section{Hit Reconstruction and Position Calibration\label{sec:highlevelcalib}}

\begin{figure}[hbt]
  \begin{center}
    \mbox{
      \subfigure[]{\scalebox{0.90}{
          \epsfig{file=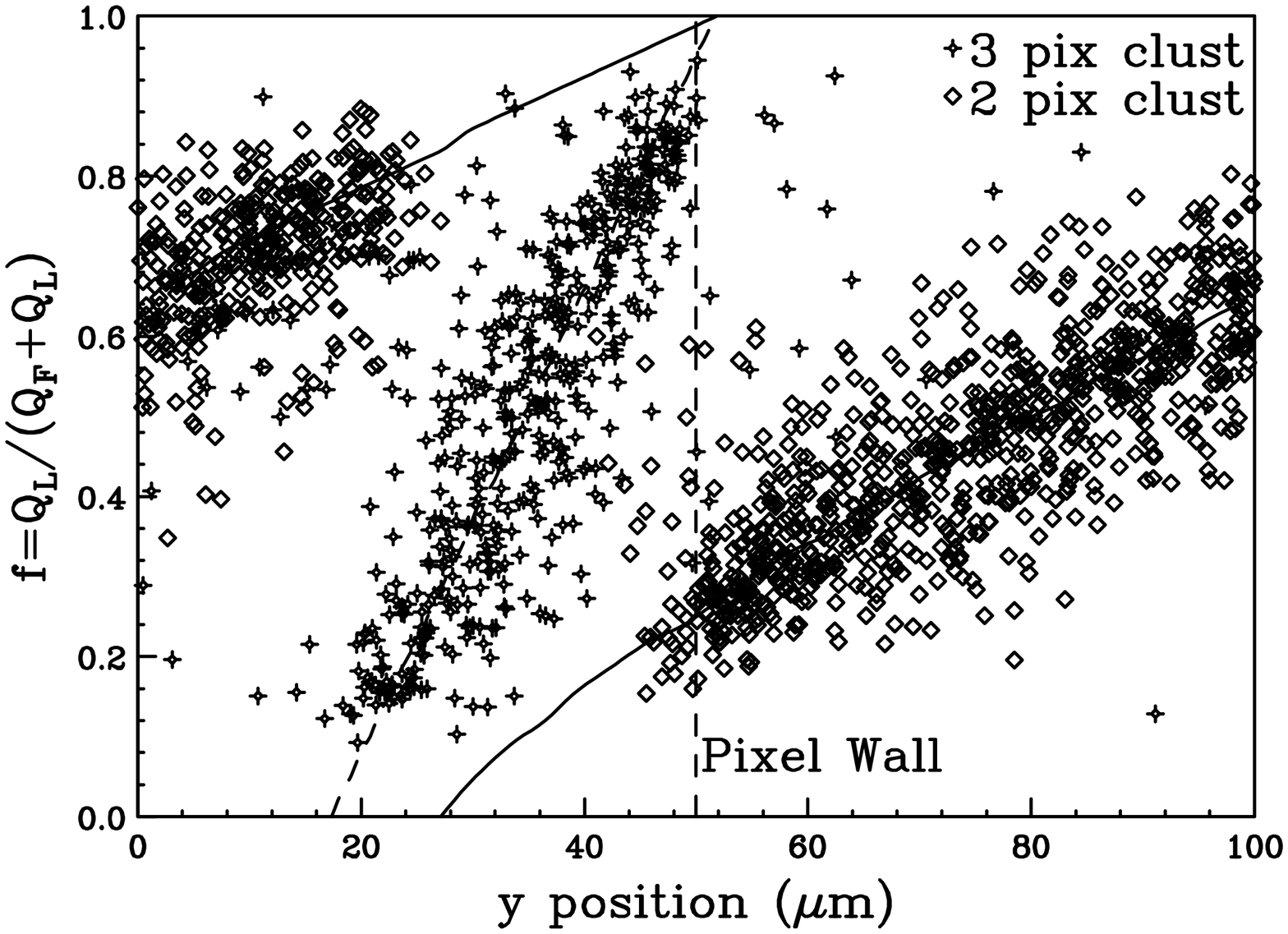,width=\linewidth}
          \label{fig:yvsf_barrel}
      }}
    }
    \mbox{
      \subfigure[]{\scalebox{0.90}{
	            \epsfig{file=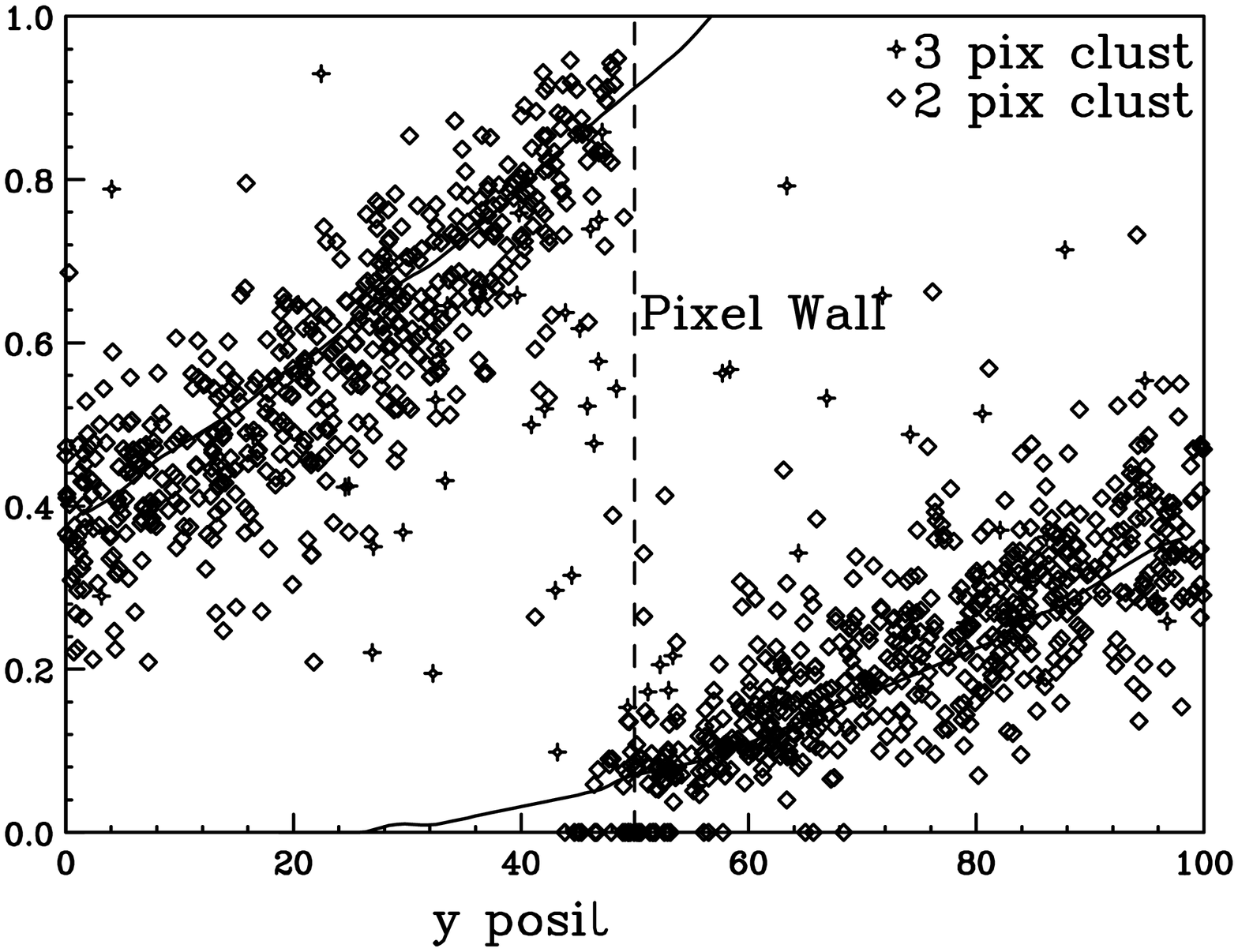,width=\linewidth}
          \label{fig:yvsf_barrel_irr}
      }}
    }
    \caption{The azimuthal charge-sharing functions for 2- and 3-pixel clusters in the CMS pixel barrel for new sensors (a) and sensors irradiated to a fluence of $\Phi = 5.9 \times 10^{14}$ n$_{\rm{eq}}$/cm$^2$ (b).  The fraction of charge found in the last pixel as compared with both end pixels is plotted as a function of the local coordinate $y$ in microns.}
  \end{center}
\end{figure}

The simulation has been used to study how the sharing of charge among neighboring pixels is affected by radiation damage.  A description of this behavior is needed to adjust the pixel hit reconstruction algorithm as the detector ages.  In what follows we refer to the reconstruction of barrel hits in the $r-\phi$ plane, where the charge drift is affected by the Lorentz deflection. Simulated tracks are perendicular to the sensor plane along the $r-\phi$ direction. Hits in the pixel detectors are reconstructed by first searching for clusters of pixels with signals above the readout threshold of 2000 electrons.  The signals are then summed along the rows and columns of the cluster to produce 1-D projected signal profiles.  In the pixel barrel, the azimuthal Lorentz drift produces 2-pixel and 3-pixel wide projected clusters in the local $y$-direction (global $\phi$-direction).  The charge fraction is defined as $f=Q_L/(Q_F+Q_L)$ where $Q_L$ is the signal of the last projected pixel in the cluster and $Q_F$ is the signal of the first projected pixel.  The charge fraction of a number of simulated hits is plotted in Fig~\ref{fig:yvsf_barrel} as a function of the track coordinate $y$ at the midplane of the sensor.  The $y$-coordinate is plotted from the center of one pixel ($y=0$) to the center of the neighboring pixel ($y=100\ \mu$m).  The two-pixel hits are plotted as open diamonds and the three-pixel hits are plotted as crosses.  These functions $y=F_2(f), F_3(f)$ are quite linear and have non-negligible width due to fluctuations in the deposited charge.  After irradiation, the introduction of trapping states dramatically increases the leakage current and produces space charge in the detector bulk.  The sign of space charge varies from negative near the n+ implant to positive near the p+ implant resulting in a doubly-peaked electric field \cite{Eremin:2002wq}.  The presence of the doubly-peaked field and the trapping of the signal carriers affect the charge sharing functions which are shown in Fig.~\ref{fig:yvsf_barrel_irr} for a fluence of $\Phi = 5.9 \times 10^{14}$ n$_{\rm{eq}}$/cm$^2$.  We note that nearly all of the 3-pixel clusters have vanished, the shape of the 2-pixel function $F_2(f)$ has become non-linear, and there are now 1-pixel clusters near the inter-pixel wall (shown as $f=0$ points).  

The large radiation-induced change in charge sharing that occurs during detector operation requires the implementation of a calibrate-able hit reconstruction algorithm.  The algorithm should be based upon calibration parameters that can be varied in time to remove biases and to ensure the use of a technique that optimizes the use of the available information.  The charge sharing functions are well described by the following expression
\begin{equation}
y=\left\lbrace\begin{array}{ll} y_c+y_2+\left(w+y_1-y_2\right)\cdot f^\alpha\ & 0<f<1 \\
y_c + \Delta \simeq y_c + \left(y_1+y_2\right)/2 & f = 0 \end{array}\right . \label{eq:Fdef}
\end{equation}
where: $y_c$ is the coordinate of the center of the first pixel in the cluster, $w$ is the pixel size, $y_1$ and $y_2$ define the $f=0$ and $f=1$ intercepts, the exponent $\alpha$ describes the linearity of the function, and the offset $\Delta$ is needed to account for the Lorentz-drift-induced asymmetry when the cluster size is one pixel. 

The performance of a reconstruction algorithm based upon equation~\ref{eq:Fdef} is shown in Table~\ref{tab:Fres} for samples of simulated events with new and irradiated sensors ($\Phi = 5.9 \times 10^{14}$ n$_{\rm{eq}}$/cm$^2$).  The means and the RMS widths of the differences between the reconstructed and true hit positions are tabulated as functions of the total cluster charge.  The large cluster charge events are likely to have energetic delta rays that spoil the resolution.  It is clearly important to develop a parameterization of the resolution for use in track reconstruction.  The optimal hit reconstruction parameters used in equation~\ref{eq:Fdef} vary considerably before and after irradiation.  In particular, the exponent $\alpha$ for the two-pixel clusters varies from 1.00 before irradiation to 0.575 after irradiation.
%
%
\begin{table*}[htb!]
\begin{center}
\begin{tabular}{|c|c| cc | cc | cc |}
\hline
 & & \multicolumn{2}{|c|}{New Sensor}  & \multicolumn{2}{|c|}{Irr. Sensor: w/o calibration}  & \multicolumn{2}{|c|}{Irr. Sensor: with calibration}  \\
Cluster Charge& Fraction & $\overline{\Delta y}$ & RMS & $\overline{\Delta y}$ & RMS & $\overline{\Delta y}$ & RMS \\
\hline
$Q/Q_\mathrm{avg}<0.7$ & 2\% & 0.1$\mu$m & 5.4$\mu$m &  26$\mu$m & 14$\mu$m&  0.0$\mu$m & 7.7$\mu$m \\
$0.7<Q/Q_\mathrm{avg}<1.0$ & 62\% & 0.1$\mu$m & 7.7$\mu$m &  29$\mu$m & 12$\mu$m &  0.3$\mu$m & 9.4$\mu$m\\
$1.0<Q/Q_\mathrm{avg}<1.5$ & 30\% & 0.4$\mu$m & 16$\mu$m &  29$\mu$m & 17$\mu$m & 0.2$\mu$m & 15$\mu$m \\
$1.5<Q/Q_\mathrm{avg}$ & 2\% & 19$\mu$m & 63$\mu$m &  34$\mu$m & 52$\mu$m &  11$\mu$m & 53$\mu$m \\ \hline
All $Q$ & 100\% & 1$\mu$m & 18$\mu$m &  29$\mu$m & 19$\mu$m &  0.6$\mu$m & 18$\mu$m \\ \hline
\end{tabular}
\end{center}
\caption{Simulated bias $\overline{\Delta y}$ and resolution (RMS) of the pixel hit reconstruction algorithm for different cluster charge bins before and after irradiation to $\Phi = 5.9 \times 10^{14}$ n$_{\rm{eq}}$/cm$^2$.  The fractions of the sample in each cluster charge bin are also listed.}
\label{tab:Fres}
\end{table*}

The parameters of the pixel hit reconstruction algorithm are sensitive to bias voltage, operating temperature, and irradiation fluence.  One can anticipate that different regions of the detector will require different parameter sets and that all will change in time.  We believe that it will be possible to use data to tune the {\sc pixelav} simulation and then use the simulation to derive the parameters to track the detector aging effects.  However, it is also possible to extract the shapes of the charge sharing functions $F(f)$ directly from data using the traditional ``$\eta$-technique''~\cite{Chiochia:TIME05}.  The technique works as follows.  It is assumed that a group of pixels is uniformly illuminated by a set of parallel tracks.  The traditional formulation uses the variable $\eta\equiv(Q_L-Q_F)/(Q_F+Q_L)=2f-1$ and with the uniform illumination assumption, it permits the extraction of $F(f)=F(\eta)$ up to an unknown integration constant $C$,
\begin{eqnarray}
&&\frac{dN}{d\eta}=\frac{dN}{dy}\frac{dy}{d\eta}=\frac{N}{w}\frac{dy}{d\eta} \nonumber \\
\to&&y=\frac{w}{N}\int^{\eta}_{-1}d\eta^\prime\frac{dN}{\d\eta^\prime}+C. \label{eq:eta_def}
\end{eqnarray}
This provides a significant constraint on $F(f)$ but still requires input from the simulation or another technique to determine $C$.  The constant $C$ is related to Lorentz-drift which has traditionally been studied by searching for an azimuthal cluster size minimum as a function of track angle.  Unfortunately, the small azimuthal angular acceptance of a pixel barrel module ($\sim\pm$10$^\circ$) and the large size of the Lorentz angle in the pixels ($\sim23^\circ$) precludes this approach. An alternative method, based on the measurement of the Lorentz deflection of long clusters was proposed in~\cite{Dorokhov:2004xk}. This method allows the measurement of the Lorentz deflection and the extraction of the electric field as function of the sensor depth.

\section{Conclusions\label{sec:conclusions}}

In this paper a detailed simulation of the silicon pixel sensors for the
CMS experiment was used to study charge sharing and the 
calibration of the position reconstruction after heavy irradiation. 
The simulation shows that position resolution can be improved after irradiation by 
parameterizing the dependence of charge sharing on the inter-pixel
position. The procedure requires the tuning of the {\sc PIXELAV} 
simulation along the operation of the CMS pixel detector, using data 
collected with colliding beams.


\bibliographystyle{elsart-num}    


\bibliography{refs}             

\end{document}